\documentstyle[preprint,aps]{revtex}

\begin{document}
\draft
\title{Lifetime Measurements in $^{120}$Xe\\}
\author{J. C. Walpe, B. F. Davis, S. Naguleswaran, W. Reviol
\thanks{Present address: University of Tennessee, Knoxville, TN 37996.},
U. Garg}
\address{Department of Physics, University of Notre Dame, Notre Dame, IN
46556.\\}
\author{Xing-Wang Pan and Da Hsuan Feng}
\address{Department of Physics, Drexel University, Philadelphia, PA 19104.\\}
\author{J. X. Saladin}
\address{Department of Physics, University of Pittsburgh, Pittsburgh, PA
15213.\\}
\date{\today}
\maketitle
\begin{abstract}
Lifetimes for the lowest three transitions in the nucleus $^{120}$Xe have
been measured using the Recoil Distance Technique. Our data indicate that
the lifetime for the $2_{1}^{+} \rightarrow 0_{1}^{+}$ transition is more
than a factor of two lower than the previously adopted value and is in
keeping with more recent measurements performed on this nucleus. The
theoretical implications of this discrepancy and the possible reason for
the erroneous earlier results are discussed. All measured lifetimes in
$^{120}$Xe, as well as the systematics of the lifetimes of the 2$_{1}^{+}$
states in Xe isotopes, are compared with predictions of various models. The
available data are best described by the Fermion Dynamic Symmetry Model
(FDSM).
\end{abstract}
\pacs{PACS numbers: 21.10.Tg, 23.20.-g, 21.60.Fw, 27.60.+j}

\narrowtext

The neutron deficient $^{120}$Xe nucleus (Z=54, N=66) lies in a
transitional region where the neutrons occupy orbitals located midway
between the closed shells of N=50 and N=82. The low-lying collective states
of the nuclei in this region appear to be described well in the O(6) limit
of the Interacting Boson Model (IBM) \cite{casten}, with a clear test of
the model being the measured lifetimes of these states. The first lifetime
measurements in this nucleus, by Kutschera {\em et al.}
at Heidelberg \cite{kut}, yielded a lifetime of $\tau$ = 124(15) ps for the
$2_{1}^{+} \rightarrow 0_{1}^{+}$ transition. Subsequent measurements made
at Notre Dame \cite{nd1} gave a value of $\tau$ = 122(26) ps, which
appeared to confirm the value of Ref.\cite{kut}.
Indeed, the extensive compilation of Raman {\em et al.} \cite{raman1},
gives $\tau$ = 122 ps as the
adopted value for the lifetime of this transition. This compilation also
shows a "saturation" effect of the
$B(E2; 0^{+} \rightarrow 2^{+})$ value for the even-even Xe isotopes at
mid-shell (around N=66). This effect has been
attributed to the filling of the neutron orbitals \cite{raman2} since
their contribution to the total quadrupole moment is very small. This
effect also has been explained in IBM-2 by incorporating Pauli spin factors
into the Hamiltonian operators \cite{otsuka}. Very recently, however, a
discrepancy has arisen over the value of the lifetimes in this nucleus, in
that Dewald {\em et al.}, in a series of measurements at K\"{o}ln, have
reported the lifetime of the $2_{1}^{+} \rightarrow 0_{1}^{+}$ transition
to be $\tau$ = 66(3.9) ps (cited in Ref.\cite{raman1}), $\tau$ = 53(2)
ps\cite{dewald1}, and $\tau$ = 75(7) ps\cite{dewald2}. These values differ
by roughly a factor of two from the earlier measurements of Refs.\cite{kut}
and \cite{nd1} and a direct consequence of the new value is the
disappearance of the aforementioned "saturation" effect.

This paper describes our attempts to resolve this discrepancy. We have
remeasured the lifetimes of states in $^{120}$Xe using the Recoil Distance
Method (RDM) \cite{alex} and our new result for the lifetime of the
$2_{1}^{+} \rightarrow 0_{1}^{+}$ transition ($\tau$ = 64(5) ps) is
consistent with the K\"{o}ln results. Further, this result is in very good
agreement with the lifetime for this state obtained in a contemporaneous
measurement by Mantica {\em et al.} who used the fast-timing coincidence
technique, with the nucleus $^{120}$Xe populated via $\beta$$^{+}$/EC decay
of $^{120}$Cs \cite{mantica}. We have also identified the most likely
reason for the earlier erroneous results. A preliminary report on these
measurements has been made previously \cite{nd2}.

The experiment was carried out at the University of Notre Dame's Nuclear
Structure Laboratory, using the $^{106}Pd(^{18}O, 4n)^{120}Xe$ reaction at a
beam energy of 70 MeV; CASCADE calculations and a brief excitation function
measurement
were employed to determine the beam energy for the optimal population of
the 4n channel. The target was
a stretched, self-supporting, $^{106}$Pd foil of 762 $\mu g/cm^{2}$
thickness. It was mounted in the Notre Dame 'plunger' device, consisting of
three dc actuators which are used for precision
placement of the target foil with respect to a stopper foil (a stretched,
self-supporting, Au foil of 2.9 $mg/cm^{2}$ thickness). The shortest
target-stopper distance, as determined through capacitance measurements,
was 18 $\mu m$.
$\gamma$-ray spectra were recorded in singles mode using four
Compton-suppressed HPGe detectors of the Pittsburgh $\gamma$-ray Array
placed at angles of 31.3$^{\circ}$,
90.2$^{\circ}$ , 146.5$^{\circ}$ , and -32.3$^{\circ}$ with respect to the
beam direction. Data were collected for approximately 4 hours at each of 24
target-to-stopper distances, ranging from the closest attainable distance
(corresponding to electrical contact) of 18 $\mu m$ to a maximum distance
of 2600 $\mu m$. This gave us an effective measurable lifetime range of
$\sim$2 ps to $\geq$1 ns.

Sample spectra for several recoil distances, taken with the detector placed
at 31.3$^{\circ}$, are shown in
Fig.~\ref{spectra}, with the corresponding recoil distance given in the
upper right corner of each spectrum. The transitions of interest are
labelled and the brackets mark the positions of the Doppler-shifted and
unshifted peaks. Lifetime information was reliably extracted for the
first three yrast transitions of $^{120}$Xe. For each transition, a set
of ratios, $R_{d}$, defined as the ratio of the unshifted $\gamma$-ray
intensity to the total intensity at recoil distance $d$, were determined.
Each set of $R_{d}$ defined a decay curve for a given transition. All such
$R_{d}$ curves were fitted with a combination of exponential functions and
the lifetime for each level was extracted from these fits. This fitting was
performed using the computer code LIFETIME \cite{wells}.
In extracting lifetime information, this code allows the following
corrections to be applied to the data: changes in the solid angle subtended
by the detectors due to the changing ion position along the flight path;
changes in solid angle subtended by the detector due to the relativistic
motion of the ion; changes in the angular distribution due to the
attenuation of alignment while the ion was in flight; and, slowing of the
ion in the stopper material. Corrections were also made to the data to
account for the detector efficiency and internal conversion. The most
significant correction made, however, was that for the effects of cascade
feeding, both observed and unobserved, from higher lying states. To account
for this, we assumed a two-step feeding process into each level for which
the observed intensity feeding into the level was less than the observed
intensity decaying out of the level. One of the feedings is from the
next highest transition in the yrast cascade, while the other represents
all unobserved feeding ({\em i.e.} feeding from the $\gamma$-ray
continuum and non-yrast states). The relative intensities for the observed
states were determined from the data collected by the detector placed at
90.2$^{\circ}$. Initial relative intensities for the levels representing
the unobserved feeding were determined by taking the difference between the
observed intensity into a given level and the observed intensity out of the
said level. In the fitting, then, both the lifetimes and the initial
relative strengths (at t = 0) of all levels representing unobserved feeding
were treated as variables to obtain the best fits to the experimental data.

The fits to the data for
the $2^{+} \rightarrow 0^{+}$, $4^{+} \rightarrow 2^{+}$ and $6^{+}
\rightarrow 4^{+}$ transitions are shown in Fig.~\ref{ratio}, and the
lifetimes (as well as the $B(E2)$ values) corresponding to
these fits are presented in Table I. Also
included in Table I, for comparison, are: the previously adopted value from
Ref.\cite{kut}, the most recent results from K\"{o}ln \cite{dewald2},
and the results from Ref.\cite{mantica}. As can be seen, our results are in
agreement (within errors) with those of Ref.\cite{dewald1,dewald2}
and in remarkably good agreement with that of Ref.\cite{mantica}.

We have investigated in some detail the difference between our present
result for the lifetime of the $2^{+}$ state and that of
Refs.\cite{kut,nd1} and can attribute it to: (a) a more accurate treatment
of the side-feeding into the ground state band; and, (b) a larger range of
recoil distances (and hence flight times) covered in the present
measurement. In effect, the two are related since the larger range of
recoil distances was crucial in identifying the side-feeding lifetimes.
Indeed, our data is best fitted by assuming a long-lived transition (of
unknown origin) feeding into the $6^{+}$ state (or higher) with an
estimated lifetime of 3500 ps. If this feeding transition were ignored in
the current analysis, and only the data corresponding to the range of
distances covered in Refs.\cite{kut} and \cite{nd1} included in the fits,
the resulting lifetime for the $2^{+} \rightarrow 0^{+}$ transition would
be $\tau$ $\sim$ 130 ps, almost identical to that of Ref.\cite{kut}.

We have calculated the B(E2) values for the first few yrast states in
$^{120}$Xe in the framework of the Fermion Dynamic Symmetry Model (FDSM)
and the calculated values also are included in Table I. The building blocks
of the FDSM are the correlated $S$ and $D$
(monopole and quadrupole) fermion pairs. This model was recently
extensively reviewed \cite{wufeng} and thus only the salient features will
be discussed here.

For the Xe isotopes, the combined neutron--proton FDSM highest symmetry is
$SO^{\nu}(8)\times SO^{\pi}(8)$, and the Hamiltonian is:
\begin{equation}
 H=-0.064S^{\dag}_{\pi}S_{\pi}^{\phantom{\dag}}
    -0.087S^{\dag}_{\nu}S_{\nu}^{\phantom{\dag}}+
    0.059 P^{2}_{\pi} \cdot P^{2}_{\pi}
    -0.010 P^{2}_{\nu} \cdot P^{2}_{\nu}
    -0.255 P^{2}_{\pi} \cdot P^{2}_{\nu} .
\end{equation}
All the operators in Eq.(1) are defined in \cite{wufeng}, and the strengths
of the interactions of Eq. (1) are in units of MeV. In computing the
$B(E2)$'s, the proton (neutron) effective charge $e_{\pi}$ ($e_{\nu}$) is
fixed at 0.19 eb (0.16 eb).

The model space is restricted to the $S$--$D$ subspace in the normal-parity
shells (heritage $u=0$, corresponding to no broken pairs).  Although the
particles in the abnormal-parity levels are not included explicitly, they
are included effectively by the constraint that there is a distribution of
particles between the normal and the abnormal parity levels. The number of
pairs $(N_1$) in the normal-parity levels is treated as a good quantum
number and is calculated from the semi-empirical formula determined
globally from the ground state spin of the odd-mass nuclei \cite{wufeng}.
For instance, according to the semi-empirical formula, $N_{\pi1}$=3 for
$^{126}$Xe and $^{128}$Xe, while $N_{\pi1}$=4 for $^{122}$Xe and
$^{124}$Xe. This difference in neutron pair number causes the staggering of
the B(E2) values with neutron number.

In Fig.~\ref{systematics}, the xenon $B(E2;0^{+} \rightarrow 2^{+})$ values
as a function of the neutron number from the present work, along with the
results reported in Ref.\cite{kut}, Ref.\cite{dewald2} and
Ref.\cite{mantica} are shown. The calculated B(E2)'s for these states using
the IBM-2 (with Pauli factors)\cite{otsuka} and the FDSM \cite{wufeng} are
also presented in the same figure. From the figure, it is quite clear that
the previously accepted experimental $B(E2; 0_{1}^{+} \rightarrow
2_{1}^{+})$ value for $^{120}$Xe and the inclusive systematic behavior of
the $B(E2; 0_{1}^{+} \rightarrow 2_{1}^{+})$ values in the even-A Xe
isotopes can be reproduced well by the IBM-2 calculations including Pauli
blocking, as was pointed out previously.  However, when the present
measurements are included, the IBM-2 calculations do show a significant
deviation from the experimental value for the N=66 ($^{120}$Xe) case and
this "deviation" appears to be contrary to the "saturation effect"
predicted by these calculations for the mid-shell nuclei. The FDSM
calculations, on the other hand, reproduce the data well, not only for the
specific case of $^{120}$Xe, but also in terms of the overall trend of the
B(E2) values for these isotopes in the mid-shell region.  It should be
pointed out, however, that in the FDSM calculations, there is a transition
from the O(5) symmetry (specifically, the O(5) scheme of a
O(6)+pairing Hamiltonian \cite{panfeng}) for the heavier Xe isotopes to the
O(6) symmetry for the lighter ones. This transition between group
structures occurs from $^{126}$Xe to $^{124}$Xe because of the change in
the particle number used in the calculations.  Intuitively, this transition
is due to the linear versus quadratic particle number dependence of the
pairing and quadrupole-quadrupole interactions, respectively.

In summary, we have remeasured the lifetimes for the first three yrast
transitions in the nucleus $^{120}$Xe, in order to investigate a factor of
two discrepancy in the measured lifetime of the lowest $2^{+} \rightarrow
0^{+}$ transition. Our results show an agreement with more recent
measurements, which are in contrast with the previously adopted lifetime
for this transition. We have determined that this difference can be
attributed to the presence of a long-lived ($\tau$ $\sim$ 3500 ps)
transition feeding into the 6$^{+}$ (or higher) state. If the present
results are taken into account, the $B(E2; 0_{1}^{+} \rightarrow
2_{1}^{+})$ values for the Xe isotopes are no longer best described by the
IBM-2 (with Pauli spin factors) model. Instead, the relatively new FDSM
seems to model
the present results rather well and, indeed, appears to best fit the
overall trend of the B(E2) values in this neutron mid-shell region.

This work has been supported in part by the National Science Foundation via
grants to the University of Notre Dame, Drexel University, and the
University of Pittsburgh.

\begin{table}
\caption{ Lifetimes and associated B(E2) values for $^{120}$Xe}
\
\begin{tabular}{cccccccc}
E(keV)&$I_{i} \rightarrow I_{f}$&$ \tau$(ps)&B(E2)($e^{2}b^{2}$)&$\tau$(ps)
\tablenotemark[1]&$\tau$(ps) \tablenotemark[2]&$\tau$(ps) \tablenotemark[3]
&B(E2) ($e^{2} b^{2}$) \tablenotemark[4]\\
\tableline
323.0&$2^{+} \rightarrow 0^{+}$&64$\pm$5&0.36$\pm$0.03&124$\pm$15&
75$\pm$7&64$\pm$4&0.36 \\
473.3&$4^{+} \rightarrow 2^{+}$&8.1$\pm$0.8&0.42$\pm$0.004&8.8$\pm$1.8&8.7
$\pm$0.9&&0.48 \\
600.9&$6^{+} \rightarrow 4^{+}$&2.5$\pm$0.3&0.44$\pm$0.005&($ < $ 5.0)&1.5/
3.0&&0.51 \\
701.7&$8^{+} \rightarrow 6^{+}$&2.6$\pm$0.3\tablenotemark[5]&($ > $ 0.18)&
0.9$\pm$0.4&&&0.48 \\
\end{tabular}
\vspace{0.25cm}
\tablenotetext[1]{ Ref. \cite{kut}.}
\tablenotetext[2]{ Ref. \cite{dewald2}.}
\tablenotetext[3]{ Ref. \cite{mantica}.}
\tablenotetext[4]{ see text.}
\tablenotetext[5]{ Lifetime of the $8^{+}$ level could not be separated
from the side feeding lifetime.  The value given is, thus, an upper limit.}
\end{table}

\begin{figure}
\caption[]{ Sample spectra from the RDM data for $^{120}$Xe at the
indicated recoil distances. The transitions of interest are marked by
brackets.}
\label{spectra}
\caption[]{ Fits to the ratios R$_{d}$ as a function of the recoil distance
for the lowest three yrast transitions in $^{120}$Xe as obtained from
LIFETIME (see text).}
\label{ratio}
\caption[]{ $B(E2;0^{+} \rightarrow 2^{+})$ values vs. neutron number for
the Xe isotopes. The calculated values from different models are also shown
superimposed.}
\label{systematics}
\end{figure}
\end{document}